\documentclass[11pt]{article}

\usepackage{natbib} 
\bibpunct{[}{]}{;}{n}{,}{,} 

\title{Experiences from Software Engineering of Large Scale AMR
  Multiphysics Code Frameworks}
\author{A. Dubey and B. Van Straalen}
\begin{document}
\maketitle

\section{Introduction}
\label{sec:intro}
Among the present generation of multiphysics HPC simulation codes
there are many that are built upon general infrastructural
frameworks. This is especially true of the codes that make use of
structured adaptive mesh refinement (SAMR) because of unique demands
placed on the housekeeping aspects of the code. They have varying
degrees of abstractions between the infrastructure such as mesh
management and IO and the numerics of the physics solvers. In this
report we summarize the experiences and lessons learned from two
of such major software efforts, FLASH \cite{Dubey2009} and Chombo
\cite{chombo2009}. 

Both Chombo and FLASH are built on top of the same SAMR
\cite{Berger1989} principles, however, their purpose and reach are
very different. Chombo is primarily an AMR framework that comes with
built in solver technologies. Typical Chombo users build application
codes for their domains by treating Chombo as a toolbox. Therefore,
Chombo has been the basis of such application codes as Bisicles,
ChomboCrunch, Compass and several others. FLASH, on the other hand, is a
complete application code that can use Chombo as one of its mesh packages. The
greater emphasis in FLASH development is on the physics modeling
capabilities. Typical FLASH users combine vast majority of capabilities
provided by the code in different ways, customize some of them and/or
add relatively small amount of code of their own. Chombo expects its
sophisticated users to modify some of its lower levels, FLASH takes a
great deal of trouble to avoid such occurrences.  Because of these
differences in approach there are differences in the designs of the
two codes, however, there are many more similarities in their software
engineering and process than there are differences. 

\section{Software Engineering}
\label{software engineering}
FLASH started its life as an amalgamation of three independent codes
written predominantly in f77 style. From the beginning the principle
purpose was to be able to simulate phenomenon of scientific interest
at the earliest possible opportunity. The architecture and modularity
was achieved by unraveling the data structures and lateral
dependencies over several iterations. Chombo, on the other hand,
started as not-backward-compatible branch of the BoxLib Framework, a
collection of tools to manage adaptive mesh. The reason for this 
bifurcation was to serve the divergent needs of two groups using
Boxlib as their basic source. The Chombo team made significant changes
to the API layers above Boxlib that were best suited for their own purposes. 
The other major difference
that influenced the design of the two codes was the language. Chombo
had the advantage of language supported object oriented features,
while FLASH imposed its own object oriented approach on
non-object-oriented code kernels through
the use of unix directory structure and a limited
domain-specific-language (DSL) for configuration. This difference is 
reflected in ways that encapsulation and abstractions play out in the
two codes.  
 
However, the maintenance process is very similar in both codes. They
use subversion for version control, have well defined coding standards
and ongoing verification practices. Both the codes  have been largely
successful in realizing the separation of concerns between the
numerical and parallel complexity of the codes through modularization
and adoption of component based architectures. The abstractions and
modularizations are usually accompanied by some reduction in the
overall performance of the code. Also, given the relatively short
shelf life of computing platforms and the large source bases that
these codes have, portability is an ongoing priority. The judicious
trade-off between maintainability, portability and performance has
been the hallmark of these and many other successful SAMR and other
multiphysics codes.  The most important area of similarity between the
codes is that they both had long term sustained funding in the initial
stages to devote to code infrastructure. Resources could be allocated
for designing the code architecture and the appropriate mechanisms for
ongoing code maintenance and verification. The current lack of such
sustained funding is a big threat to these and other large codes in
similar circumstances. 

\section{Lessons Learned}
\label{lessons}
The following are the combined lessons learned by the two code teams
that might be of interest to other projects at similar scales.

\noindent{\bf Public Releases}: Early and frequent public releases of the code
  are greatly beneficial to the overall code quality. The general
  tendency to keep the code private for perceived advantage over the
  scientific competition is misplaced and often detrimental to both
  code and science. Early and relatively frequent releases serve
  the dual purpose of ensuring that the coding standards are followed
  more carefully and that code verification is broader because of
  being exercised in many different ways by diverse users. Public
  releases also facilitate the reproducibility of science results and
  therefore confidence in those results.

\noindent {\bf Interdisciplinary Team}: A team with a breadth of knowledge and
expertise is compulsory almost by the definition of the work
involved. There must be domain experts as well as applied
mathematicians and software engineers in the team. At least a couple
of team members should have cross-cutting expertise. This is useful
not only in terms of understanding how the various components work
together in the software, but the presence of such individuals also
fosters trust and co-operation among the diverse team members.  
A team with broad and cross-cutting expertise is also able to better
absorb the loss of individuals with specific expertise. 

\noindent {\bf Documentation}: Extensive documentation is critical not
  only for the user population, but also internally for code
  maintenance. In addition to standard documentation such as a user's
  guide and other online resources targeted at the users, extensive
  inline documentation is critical for maintaining any code section
  that has even moderately complex logic. A well documented code
  section can be maintained by non-experts with general know-how of
  the code if necessary. 

\noindent {\bf Backward Compatibility}
Our experience has been that backward compatibility is not always
desirable during major version changes. It can get unnecessarily
expensive in terms of developer time and can clutter the code
architecture. There are several precedents for breaking
backward compatibility to the advantage of the longevity of the
code. This is especially true if a significant API change is desired,
and if the branch has adequate support to be a live project in its own
right. Caution should be exercised in doing this for under-resourced
branches because they may end up as nothing more than software
research. That is not a bad thing, but it should be expected and planned
for.  

\noindent {\bf User Support} Having a well defined user support policy is
  extremely important in convincing the community to use the
  code. While it may take significant resources initially to respond
  to all user queries, in time that effort reduces because the
  community becomes self-supporting. Providing comprehensive
  documentation which is easy to access and reference also helps
  reduce the demands for user support. 

\noindent {\bf Code Infrastructure} The least cluttered code architecture
provides the greatest flexibility and longevity. While it may seem
like stating the obvious, scientific codes very often fail to achieve
this for several reasons. The most common reason 
is feature creep: almost all codes have features that should have been
pruned because they either did not prove to be useful, or outlived
their usefulness.  Another common reason is that the code
infrastructure is often erroneously assumed to be less important than capability
development, especially when resource 
allocation is driven by near term scientific
goals.  The most important reason, however, is that it is extremely hard to achieve
architectural simplicity in a complex software with many moving
parts. It requires substantial investment on the part of the
developers to understand the requirements, the limitations, and the
idiosyncrasies of the core solvers as they relate to the infrastructure. 
The best time to devise and formalize code architecture is after the
knowledge about the core solvers has been thoroughly internalized, and
that requires a willingness to redesign and rewrite large chunks of
infrastructure code at least in the early iterations.

\noindent{Verification}: Code verification in general gets some attention
from scientific code developers, but ongoing daily testing usually
does not. Any large development effort cannot have confidence in the
code without such testing. Developers cannot comprehensively test for
unintended side effects from code modifications manually when the code
has many interoperating components. When there is daily testing with
sufficient code coverage, chances of early detection of unintended side
effects increases. The amount of effort needed to fix such inadvertently
introduced faults is higher the longer they stay undetected. 

\section{Future Sustainability}
\label{sustainability}
In the era of cluster computing with fat nodes, distributed memory computing provided a near
ideal programming model where these sometimes conflicting requirements
of performance, portability and maintainability
could be balanced. The overheads of communication primitives and bulk
synchronizations could be amortized over large computational units to
the point where they did not significantly compromise
performance. Also, since the node architectures were mostly homogeneous
even across vendors, general algorithmic or data structure
optimizations provided benefits across the board. Therefore, although
individual codes followed different paths, and differ significantly in
their details, they have arrived at remarkably similar solutions
conceptually. They are in production in multiple disciplines, are
prolific in producing scientific results, and are likely to continue
to do so for the next couple of years. This state of affairs combined
with the uncertainty in the HPC landscape at present has induced
hesitancy in prioritizing the infrastructure building in many code
projects and their funding agencies. We believe that this
could potentially be a recipe for disaster as more heterogeneous and
less reliable machines come online. 

The most frequently given argument that the code developers will
rewrite their codes for the target platforms is a valid one for
software with relatively small code bases. When the algorithms are well
understood, and refactoring the code is likely to take only a few
person-months there is nothing to be gained by anticipating trouble
and preparing for it ahead of time. However, large code bases do not
have this luxury. Some of them are likely to take several person-years
to transform the whole code base. Our experience indicates that
starting from scratch to write a whole new code is worse because of
higher number of degrees of freedom for error introduction. It is our
position that redesign and re-architecting of code frameworks does not
need to wait for the arrival of manycore and heterogeneous platforms,
because a consensus is beginning to emerge about the conceptual design
requirements, enough to enable the redesign phase to start.

The emerging paradigms for taking the codes to the next generation include
automatic code transformation and asynchronous runtime
management. While the individual tools and compilers for providing
these functionality will be different, the applications will have to provide
footholds for the related abstractions. The applications can do so by
taking the separation of concerns a step further than that between
numerical and parallel complexity. They have to articulate the
dependencies within the code much more explicitly, leave data-staging
and therefore assembly to the infrastructure and expose minimum
computation units, both spatial and temporal, that can be exploited by
the code transformation tools through fusion. 

Because of the above considerations, the refactorization and
transformations are needed at the more fundamental implementation
design level in the application codes and/or their infrastructure. The
data layout, the wrapper layers and the communication channels between
different code components have to be designed with an awareness of the
semantics of the programming abstractions using asynchronous task
management and code transformations. It is imperative that support is
provided for refactoring of the mature codes in this manner because
the changes to the architecture of most codes will be highly
disruptive and therefore labor intensive. The alternative, code
development from scratch, might succeed in a few instances, but is
unlikely to meet with broad success. The reasons are: (1) an
unconstrained design space has a potential to not converge, as
happened to many high level frameworks 12-15 years ago, (2) code
verification during refactoring remains tractable when solutions can
be compared against a known set of solvers, and (3) to build a robust
multiphysics code is long and arduous process, most mature codes have
taken 5-8 years to arrive at the level of confidence that they now
enjoy. We believe a judicious combination of disruptive and
incremental changes are the optimal way to continue to serve the cause
of science.
  


\end{document}